\documentclass[reprint,floats,floatfix,amsmath,amsfonts,amssymb,nofootinbib,aps,prd]{revtex4-2}
\usepackage{graphicx}
\usepackage{natbib}
\usepackage[T1]{fontenc}
\usepackage[utf8]{inputenc}

\newcommand{\p}{\partial}

\renewcommand{\a}{\approx}
\newcommand{\bs}{\mathbf}

\newcommand{\x}{|\bs x- \bs x'|}
\newcommand{\xx}{|\bs x- \bs x''|}

\newcommand{\s}{\scriptscriptstyle}

\newcommand{\wn}{WN-PPN}

\newcommand{\vp}{\textsf{v}} 
\newcommand{\ba}{{\alpha_{\rm e}}}
\newcommand{\bg}{{\gamma_{\rm e}}}
\newcommand{\gbg}{\gamma_{\s \Sigma}}
\newcommand{\mnras}{Mon.Not.Roy.Astron.Soc.}

\usepackage[hyperindex,colorlinks,allcolors=blue]{hyperref}

\begin{document}

\title{Post-Newtonian  $\gamma$-like parameters and the gravitational slip in scalar-tensor and $f(R)$ theories}

\author{J\'unior D. Toniato} \email{junior.toniato@ufop.edu.br}
\affiliation{Departamento de F\'isica, Intituto de Ciências Exatas e Biológicas,\\
Universidade Federal de Ouro Preto, 35400-000, Ouro Preto\;-\;MG, Brazil}

\author{Davi C. Rodrigues} \email{davi.rodrigues@cosmo-ufes.org}
\affiliation{N\'ucleo de Astrof\'isica e Cosmologia (Cosmo-Ufes)   \& Departamento de Física,  Universidade Federal do Esp\'irito Santo, 29075-910, Vitória\;-\;ES,  Brazil.}

\date{\today}

\begin{abstract}

We review the fundamentals and highlight the differences between some commonly used definitions for the PPN gamma parameter ($\gamma$) and the gravitational slip ($\eta$). Here we stress the usefulness of a gamma-like parameter used by Berry and Gair ($\gbg$) that parametrizes the bending of light and the Shapiro time delay in situations in which the standard $\gamma$ cannot be promptly used. First we apply our considerations to two well known cases, but for which some conflicting results can be found: massive Brans-Dicke gravity and $f(R)$ gravity (both the metric and the Palatini versions). Although the slip parameter is always well defined, it has in general no direct relation to either light deflection or the Shapiro time delay, hence care should be taken on imposing the PPN $\gamma$ bounds on the slip.  We stress that, for any system with a well posed Newtonian limit, Palatini $f(R)$ theories always have $\gamma = 1$; while metric $f(R)$ theories can only have two values: either 1 or 1/2. The extension towards Horndeski gravity shows no qualitative surprises, and $\gbg$ is a constant in this context (only assuming that the Horndeski potentials can be approximated by analytical functions). This implies that a precise study on the bending of light for different impact parameters can in principle be used to rule out the complete Horndeski action as an action for gravity. Also, we comment on the consequences for $\gamma$ inferences at external galaxies.
\end{abstract}

\maketitle

\section{Introduction}

Parameterized Post-Newtonian (PPN) formalisms \cite{Eddington:1922b, Robertson:1962rc, Schiff:1967, Will:1972zz, Damour:1992we, Will:1993ns, Will:2014kxa} provide  practical and rigorous procedures to infer model constraints from several different observations.  Although there are some relevant differences between different PPN formalisms, they all require the model to be expressed in a precise way and to satisfy certain set of hypothesis. If these conditions are met, the PPN parameters of the model are found and they can be promptly compared to observational bounds.  There is currently large interest in using PPN parameters beyond the classical domain of the solar system (e.g., \cite{Bolton:2006yz, Clifton:2018cef, Collett:2018gpf}), which is welcome to test both general relativity (GR) and newer proposals for gravity. Other parametrizations, related or inspired by the PPN formalisms, have also emerged (e.g., \cite{Bertschinger:2006aw, Amendola:2007rr, Daniel:2009kr, Baker:2012zs, Sanghai:2016tbi}). We stress here certain subtle differences which may, under some circumstances, be important and change the inferred parameter bounds. Although our conclusions can be applied in different contexts (including the solar system), our main physical motivation comes from constraining gravitational parameters by light deflection from external galaxies (as in Refs.~\cite{Bolton:2006yz, Cao:2017nnq, Collett:2018gpf}). In this context, the  Newtonian limit is commonly expected, but it may not hold, hence considering both the cases is valuable.

We start by reviewing the fundamentals of the standard PPN formalism, due to Will and Nordtvedt and referred as \wn, with emphasis to its $\gamma$ parameter \cite{Will:1993ns, Will:2014kxa}. The parameter $\gamma$ has a special role since it is the single post-Newtonian (PN) parameter that appears in a first order metric expansion and it is directly associated with  tests concerning the propagation of electromagnetic waves. Based on that, we will explore  two others $\gamma$-like parameters (here denoted by $\bg$ and $\gbg$) that appears within an extension of the \wn~parametrization, addressing the similarities and differences between the gammas and the gravitational slip parameter ($\eta$). We apply these discussions to the case of scalar-tensor and $f(R)$ theories. We understand that there are some common misconceptions, in particular on the view that the slip can be seen as an effective gamma, which can lead to wrong interpretations on the behaviour of electromagnetic waves and the numerical bounds of the theory.

The paper is organized as follows. In Section \ref{sec:ppn} we review the fundamentals of \wn~formalism and the physical meaning of the gamma parameter, also, following \cite{Berry:2011pb}, we introduce $\gbg$ as a more general parametrization for the dynamics of electromagnetic waves. By the end of Section \ref{sec:ppn}, we  start to explain some of the relevant differences between $\eta$, $\gamma$ and $\gbg$.  Section \ref{sec:st} describes the identification of the slip and gamma parameters in generalized Brans-Dicke theories and  $f(R)$ theories. Section \ref{sec:horndeski} extends the previous arguments towards  Horndeski theory. Finally, we present our conclusions in Section \ref{sec:conclusion}.


\section{Gamma parameters and gravitational slips}\label{sec:ppn}

\subsection{WN-PPN introduction}
The large number of different physical constraints that can be extracted from solar system tests lead to the development of  practical mechanisms to confront a theory's prediction with observational data. Eddington was one of the pioneers \cite{Eddington:1922b}, followed by Robertson and Schiff \cite{Robertson:1962rc,Schiff:1967}. The Eddington-Robertson-Schiff (ERS) formalism  describes the planets as test particles moving along geodesics of a spherical and static background that extends the corresponding GR solution by introducing two parameters, usually denoted by $\gamma$ and $\beta$. A more recent and well known PPN formalism is based on works by Will and Nordtvedt (\wn), where  a continuous matter description of celestial bodies is used \cite{Will:1972zz,Will:1993ns}. The \wn~formalism uses ten metric parameters, with nine of them being directly constrained by observations \cite{Will:2014kxa}. 

One of the advantages of the WN-PPN formalism resides in the fact that, once the PN metric of a theoretical model is obtained, one just needs to read off the PPN parameters from the metric in order to confront theory and observation. Contrary to the ERS case, it is not necessary to find explicit metric solutions (or assume spherical symmetry), only the relation between the metric and certain functionals need to be found. All the study on the equations of motion was already performed using the general metric of the formalism. However, there are limitations since several of the alternative theories cannot be parametrized according to the original \wn~scheme. One thus need to evaluate how the new terms in the metric expansion can affect the equations of motion in order to infer observational constraints in any free parameter of the theory (see e.g., \cite{Toniato:2017wmk}).

Extensions of GR that introduce a scalar field $\phi$ are among the most frequently considered. Scalar-tensor models that depend on a potential $V(\phi)$  cannot be in general be studied in \wn, or even in a more general PN context. This since the potential may spoil the existence of a Newtonian gravity limit, while PN bounds on $\gamma$, for instance, requires the existence of a Newtonian limit up to certain precise order. We will revist in detail this point in this work.

\subsection{The standard gamma definition}\label{sec:gamma}

We start by defining the parameter $\gamma$ in accordance with the WN-PPN approach \cite{Will:1993ns,Will:2014kxa}. The latter requires that the following assumptions must hold up to the expansion order being considered:
\renewcommand{\labelenumi}{\textit{\roman{enumi}})}
\begin{enumerate}
\item The matter of the system can be described as a perfect fluid.
\item The relevant spacetime for the system is asymptotically flat.\footnote{At distance scales much larger than that of the system,  asymptotically flatness need not to be satisfied.}
\item The metric can be expanded about Minkowski, where the $n$-th order expansion is of the same order of $(v/c)^n$.\footnote{Where $v$ is the fluid velocity and $c$ is the speed of light.} 
\item A well defined Newtonian limit must exist. 
\end{enumerate}

The WN-PPN formalism, in its original form, considers that metric perturbations depends on 10 functionals of the fluid variables (the PPN potentials) and 10 dimensionless constants that can be constrained from experimental and observational bounds (the PPN parameters). The functionals were chosen with considerations on fairness and simplicity, being such that GR is included as a special case (further details can be found in Sec.~4.1 from Ref.~\cite{Will:1993ns}). Therefore, not all theories can be written in that form. If a theory can be shown to be a particular case of that specific metric parametrization, one simply has to consult a table with the bounds on each of the constant coefficients. This procedure is used in Ref. \cite{Toniato:2017wmk}, for example. If the theory depends on functionals that are not among the 10 original ones, it is in general necessary to proceed to the full PN equations of motion to find the physical bounds, as it was done, for instance, in Ref.~\cite{Toniato:2019rrd}.

In order to compare to cosmological perturbative parametrizations (which is the context in which the slip is commonly defined), we  only need the WN-PPN metric restricted to the first order metric perturbations. This restriction is sufficient to include both the Newtonian dynamics and the geometry input that determines light trajectories. In this case, the WN-PPN metric depends on a single parameter and it reads, in the PPN gauge, 
\begin{subequations} \label{ppnmetric}
\begin{align}
	&g_{00}^{\mbox{\tiny PPN}}=-1+2U + O(4), \label{ppng00}\\[.2cm]  
	&g_{0i}^{\mbox{\tiny PPN}}=0 + O(3), \label{ppng0i} \\[.2cm]
	&g_{ij}^{\mbox{\tiny PPN}}=\delta_{ij} +2\gamma U \delta_{ij} + O(4). \label{ppngij}
\end{align}
\end{subequations}
where
\begin{equation}\label{u}
U(x, t) \equiv \int  \frac{\rho(x',t)}{\x}\,d^{3}x' 
\end{equation}
is the (negative of the) Newtonian potential and $\rho$ is the rest mass density. In the above, and hereafter, we use units such that $G=c=1$ and the notation $O(N)$ to indicate terms of order $v^N$.  For a bounded Newtonian system, from the virial theorem, $U \sim v^2\sim O(2)$. 

The PN corrections to light dynamics are entirely determined by the second-order metric components \cite{PoissonWill}.  

The PPN approach of Damour and Esposito-Farese \cite{Damour:1992we}  is different from the \wn\,  one, but their metric  agrees with the metric \eqref{ppnmetric} up to O(2) terms.

The $O(2)$ term in $g_{00}$ (i.e., $2 U$) provides the Newtonian-order effects (with $G=c=1$). Since PPN is a post-Newtonian parametrization (as implied by its name), it is not surprising that Newtonian dynamics are already taken for granted at the appropriate expansion order. Applying PPN parametrizations on theories without a Newtonian limit is not impossible, but cannot be done blindly: it requires a ``modified PPN'' approach in which one has to return to investigate the physical meaning of the PPN parameters in the new context (e.g., \cite{Alsing:2011er}). In order to review and better detail the relevance of $\gbg$ \cite{Berry:2011pb}, even for the cases without a Newtonian limit, we will consider the equations of motion for light.

The $g_{ij}$ components include the single contribution that is PN up to the first-order metric perturbation.  Within GR it has only a single term up to $O(2)$, which is proportional to $U \delta_{ij}$. The WN-PPN formalism generalize the latter dependence by appending an arbitrary constant $\gamma$ factor. If any parameter should be called $\gamma_{\rm \s PPN}$, we understand that it is this one.  For convenience, we  call it $\gamma$ and use other symbols for other similar quantities.

One may consider other possible contributions at $O(2)$ to $g_{ij}$, however there are not many options that are in agreement with the \wn\, gauge and respect the PPN expansion. For instance, another natural candidate would be $U_{ij}$ \cite{Will:1993ns}, where
\begin{equation}
	U_{ij}(x,t) \equiv \int \frac{\rho(x',t) (x_i - x'_i)(x_j - x'_j)}{|\mathbf{x} - \mathbf{x}'|^3}d^3x'	\, .
\end{equation}
This non-diagonal quantity asymptotically decays, it has the right dimensions and it is indeed a $O(2)$ quantity. However, $U_{ij}$  can be removed by a gauge choice, being absent from the metric in the standard \wn\, gauge. 

In conclusion, up to the $O(2)$, the single PPN parameter is $\gamma$, which is a constant. This parameter is more than an arbitrary number in the metric, it has well defined physical implications. Indeed, its value changes the trajectories of photons, as explained in the next section.

\subsection{The physical meaning of $\gamma$} 
Consider the propagation of photons in the  metric expansion \eqref{ppnmetric}. The geodesic equation reads
\begin{equation}
	\frac{d^2 x^\mu}{d \lambda^2} = - \Gamma^{\mu}_{\alpha \beta} \frac{dx^\alpha}{d\lambda}\frac{dx^\beta}{d\lambda} \, ,
\end{equation}
where $\lambda$ is an affine parameter. It is useful to express the above without reference to $\lambda$. For massless particles it can be written as,
\begin{equation}\label{geod}
\dfrac{d\textsf{v}^{\mu}}{dt}=-\left(\Gamma^\mu_{\alpha\beta}-\vp^{\mu}\Gamma^{0}_{\alpha\beta}\right)\vp^{\alpha}\vp^{\beta}.
\end{equation}
where $\Gamma^{\mu}_{\alpha\beta}$ is the Levi-Civita connection and $\vp^{\mu}=dx^{\mu}/dt=(1,\vec{\vp})$ is the photon four-velocity, which satisfies, 
\begin{equation}\label{norm1}
g_{\alpha\beta}\vp^{\alpha}\vp^{\beta}=0 .
\end{equation}
Up to $O(2)$ terms, Eq.~\eqref{norm1} leads to
\begin{equation}
	\vp^i \vp^j \delta_{ij} = 1-2 (1+\gamma)  U\, 
\end{equation}
and hence,
\begin{equation}
	\vp^i=[1-(1+\gamma)U]n^i \label{norm}\, ,
\end{equation}
where $n^i$ is a normalized vector  such that $n^i n^j \delta_{ij} = 1$. The velocity  $\vp$  is light-like, hence it is not small, but $v$ is small.

From Eqs.~\eqref{geod} and \eqref{norm}, and up to $O(2)$ contributions \cite{PoissonWill},
\begin{equation}
\dfrac{dn^i}{dt}= (1+\gamma)(\delta^{ij}-n^in^j)\p_j U \, , \label{acc}
\end{equation}
where it was used that $dU/dt=\p_tU + \vp^j\p_jU\a \vp^j\p_jU$. This since time derivatives increase the perturbation order by one \cite{Will:1993ns}.

The relevant equations for the deflection of light by a static body, as well as the Shapiro time-delay effect, are directly obtained from Eqs.~\eqref{norm} and \eqref{acc} (see \cite{Will:1993ns} for further details). Both phenomena can be used to put bounds on $\gamma$. Currently, the time-delay in the propagation of light is the phenomenon responsible for the  strongest constraint, yielding \cite{Will:2014kxa},
 \begin{equation}\label{bound}
 	|\gamma-1|\lesssim 2.3\times 10^{-5}.
 \end{equation}

The parameter $\gamma$ also influences other PN effects, like the perihelion shift. However, it is the single PPN parameter to have an effect on the time-delay and deflection of light.

The bound given in \eqref{bound}, like any other bound on $\gamma$ within the WN-PPN, depends on the validity of all the \wn \, hypothesis. Among them, we recall that  only the Newtonian potential is present at the $O(2)$ order and that $\gamma$ is a constant. Scalar-tensor theories need not  to satisfy these conditions.

In the following subsection, we present an extension of the metric parametrization \eqref{ppnmetric} and a gamma parameter ($\gbg$) that captures the physical essence of $\gamma$ in this extended parametrization. 

\subsection{Extended PPN metric parametrization and $\gbg$}\label{sec:gbg}

Here we consider the PN equations of motion for photons within an extended metric with respect to Eq.~\eqref{ppnmetric}, namely

\begin{subequations} \label{barmetric}
\begin{align}
&\bar{g}_{00}=-1+2 \ba U + O(4), \label{barg00}\\[.2cm]  
&\bar{g}_{0i}=0 + O(3), \label{barg0i} \\[.2cm]
&\bar{g}_{ij}=\delta_{ij} +2 {\bg} U \delta_{ij} + O(4). \label{bargij}
\end{align}
\end{subequations}
In the above, $\bg$ and $\ba$ are arbitrary functions of space and time and we use the subscript $e$ for extended. Due to the arbitrariness on $\bg$ and $\ba$, the $U$ that appears above is just a useful notation convention to compare with the standard PPN metric results. The metric above depends on two arbitrary scalars. We remark that although $\alpha$ can be eliminated by a unit redefinition, thus without physical impact, $\ba$ can at most be eliminated locally. The same issue can be found in some other PPN extensions that consider dynamics at scales much larger than that of the solar system (e.g., \cite{Sanghai:2016tbi}).

By repeating the same computations of Eqs.~\eqref{norm} and \eqref{acc}, one then finds  
\begin{align}
&\vp^i=\big[1-\big( \ba + \bg\big)U\big]n^i, \label{normbar}\\
&\dfrac{dn^i}{dt}= (\delta^{ij}-n^in^j)\partial_j[(\ba+ \bg)U]\, \label{accbar}.
\end{align}
In the above expressions, time derivatives were once again neglected, since they are one order higher with respect to spatial variations.
From the above, one sees that the equations of motion with $\ba + \bg = 2$ are the same equations of GR for light. This holds  even if $\bg \not \approx 1$, or even if $\ba$ and $\bg$  are spacetime functions that change considerably locally. 

A direct comparison between Eqs.~\eqref{normbar}-\eqref{accbar} and \eqref{norm}-\eqref{acc}, allows one to identify an effective gamma parameter as a function of the sum $\ba + \bg$. Indeed, this observation was done by Berry and Gair  in the specific context of  $f(R)$ theories \cite{Berry:2011pb}. Let this effective gamma be denoted by  $\gbg$, with ``$\Sigma$'' as a reference for sum, and with
\begin{equation}
	\gbg \equiv  \ba + \bg - 1. \label{gammaBG}
\end{equation}

The gamma parameter above is a natural definition since: $i$)  if both $\bg$ and $\ba$ are constants, one can redefine $\rho$ such that $\ba = 1$ and $\gbg = \gamma$; $ii$) whenever $\gbg$ is a constant,  it has exactly the same role that $\gamma$ has for light propagation \cite{Will:2014kxa}. This implies, in particular, that the impact parameter dependence, associated to light bending and the Shapiro time-delay, will be the same of GR. This however does not imply that the $\gamma$ bounds can  be immediately applied to $\gbg$, since $\gamma$ is measured within a valid Newtonian limit, which is important for the gravitational mass definition, while $\gbg$ can be computed even without a Newtonian limit (hence, the meaning of the gravitational mass may not be fixed).  For instance, the important prediction of GR about the deflection of light by the Sun was only a prediction since it was known what was the (gravitational) mass of the Sun. On the other hand, even without the knowledge of the mass of the Sun, one could test how the deflection angle changes for different impact parameter values.  To be more explicit, we state the expression for light deflection as a function of a constant $\gbg$ (which are the same of \wn\, with $\gamma \to \gbg$) \cite{Will:1993ns}. For simplicity, let $\rho = M \delta^3(\mathbf r)$, hence
\begin{equation} \label{deltatheta}
	\delta\theta=\left(\frac{1+\gbg}{2}\right)\frac{4M}{d}\left(\frac{1+\cos\theta_0}{2}\right) \, .
\end{equation}
In the above, $\theta_0$ is the unperturbed (true) angle between a massive body of mass $M$ and a luminous source (e.g., the Sun and a distant star), $\delta \theta$ is the deflection angle due to the massive body and $d$ is the light impact parameter. The mass $M$ enters in the above due to $U = M /r$.


In general, for systems that can be described by the metric \eqref{barmetric}, spacetime can be fully described by the pair of functions $(\ba, \bg)$, or equivalently by $(\ba, \gbg)$. This is similar to cosmological approaches within first order perturbations (e.g.,\cite{Amendola:2007rr, Skara:2019usd}). However, our focus here is on approaches closer related to PPN. As it will be shown, for scalar-tensor theories in the Horndeski family and  $f(R)$ theories, $\gbg$ is a constant, even though $\ba$ and $\bg$ are not constants in general. Thus implying that the relation \eqref{deltatheta} is valid for this family of theories.

\subsection{Gravitational slip}\label{sec:slip}
Many efforts have been concentrated in the development of a PPN-like formalism applicable to cosmology. Bertschinger drew attention to the importance to find ways to constraint the difference between the two scalars that appear in cosmological first order perturbations  in the Newtonian gauge \cite{Bertschinger:2006aw} (which are commonly denoted by $\phi$ and $\psi$). He suggested that a comparison between lensing effects generated by a given system with the non-relativistic internal motion of the same system could be used to test the difference between $\phi$ and $\psi$. Different variations, that follow this physical setup, can be found (e.g., Ref.~\cite{Daniel:2009kr} and  references therein). 

In a cosmological context, let
\begin{equation}
ds^2 = a^2(\tau) \left [- (1 - 2 \psi) d\tau^2  +  (1 + 2 \phi)  d\mathbf{x}^2 \right]	,\label{frw}
\end{equation}
where we use  $\tau$ for the conformal time.

The gravitational slip can be defined in several different ways.  A common definition is given by the ratio of the Fourier transforms of the two scalar perturbations in the quasi-static limit \cite{Daniel:2009kr, Amendola:2012ys, Aghanim:2018eyx, Ishak:2018his}, that is
\begin{equation}
	\eta_{\mathbf{k}}  \equiv \frac{\phi_{\mathbf k}}{\psi_{\mathbf k}}.	 \label{slipk}
\end{equation}
In the above, the dependence on $\tau$ is implicit. Sometimes one also considers the slip parameter in physical space, $\eta$ (e.g., \cite{Linder:2018jil, Pizzuti:2019wte}). As it is customary, we use the same letter $\eta$ to designate it, but it is actually an independent function: it is not the Fourier transform of $\eta_{\mathbf{k}}$. The slip in physical space is defined as
\begin{equation}
	  \eta  \equiv \frac{\phi}{\psi}.	\label{slipdeff}
\end{equation}

In this work, unless otherwise specified, we use the space-dependent definition of the slip \eqref{slipdeff}, which is simply denoted by $\eta$.
We recall here that, within GR up to first order perturbations, one finds $\eta = 1$  whenever $T_{i j} = 0$ for $i\not=j$.

\subsection{Slip is not gamma} \label{sec:slipIsNotGamma}

The slip was introduced in a cosmological context, but it can be promptly particularized for the solar system one. Indeed, for the  \wn~metric \eqref{ppnmetric}, $\eta$ is a constant and it is given by 
\begin{equation}
	\eta|_{\rm \s PPN} = \gamma \,. \label{etawnppn}
\end{equation}
The subscript ``PPN'' is a reference to the metric \eqref{ppnmetric}. In particular, this equality implies that any bounds on the $\gamma$ parameter are also valid to $\eta|_{\mbox{\rm \tiny PPN}}$. Probably this relation had a role as the source of the confusion between $\gamma$ and $\eta$. The point to be stressed is that the equality $\eta = \gamma$ is wrong in general, but it holds for the metric \eqref{ppnmetric}. 

For the extended PPN-like metric \eqref{barmetric}, here labeled as ``EPPN'', one has
\begin{equation}\label{slip-bar}
	\eta|_{\s \rm EPPN}=\frac{\bg}{\ba}\, .
\end{equation}
The gravitational slip above is neither equivalent to $\gamma$, $\bg$ or $\gbg$. Hence, in general, there is no  reason for $\eta$ to satisfy any bounds that are valid for $\gamma$.  

The  physical phenomena that $\eta$  probes is not in general the same phenomena tested by $\gamma$, $\bg$ or  $\gbg$: they happen to coincide for the particular metric \eqref{ppnmetric}, but $\eta$ is commonly used with the metric \eqref{frw}. A key difference is that the parameter $\gamma$ is defined in a PN context, hence Newtonian gravity is assumed to hold up to a given order, and it is in this context that the $\gamma$ bounds are derived.  On the other hand, $\eta$ may be used in different contexts, independently on the Newtonian limit and  on the metric \eqref{ppnmetric}.

Apart from the differences mentioned above, it is instructive to consider if there is some approximation, with $\ba \not=1$, in which $\eta$ could be seen as an effective $\gamma$, in the sense of a parametrization  which could be subjected to the same observational bounds of $\gamma$. The simplest case is the following: if it is  assumed that $\ba$ can be approximated by a constant inside the  system being considered, then it is possible to redefine $\rho$ (or the gravitational constant) inside $U$ such that $\ba U \to U$, implying that $\bg U \to \eta U$. It is further needed that $\eta$ can be approximated by a constant. For this simple case, it is correct to apply the $\gamma$ bounds over $\eta$, but care should be taken to ensure that $\ba$ and $\eta$ are indeed sufficiently close to constants. Considering all the solar system tests commonly probed in the PPN context, these quantities need to be constants for a significantly large range of distances: from the largest distances of the planets orbits $\sim 30 \mbox{ au}$ (if not beyond) to the solar radius $\sim 5 \times 10^{-3} \mbox{ au}$ (since the bending of light and the Shapiro time delay are particularly susceptible and tested when light passes close to the solar radius).

It is possible to alleviate the assumptions above on  $\ba$ and $\eta$ being exact constants. Indeed, according to eqs.~\eqref{acc} and \eqref{accbar},  it is not  necessary that $\ba$ and $\eta$  are constants for parameterizing light bending, but at least that 
\begin{equation}
	|\partial_j (\ba + \ba \eta) U| \ll	|(\ba  + \ba \eta) \partial_j U| \, . \label{constantEta}
\end{equation}
If the latter holds, then  $|\ba(1+ \eta) -2|$ will be subjected to the same bounds that $|\gamma-1|$ is subjected.\footnote{Considering the theory presented in Sec.~\ref{sec:gbg}, this should be no surprise.}  That is, if  $|\gamma - 1|< B_\gamma$, where $B_\gamma$ is some  small positive number, then 
\begin{equation}  \label{alphaeetaBound}
	|\ba(1+ \eta) -2| < B_\gamma \, .
\end{equation}
Clearly, this bound depends on both $\ba$ and $\eta$, it is not a bound for $\eta$ alone. If one assumes that $|\eta-1| \lesssim B_\gamma$  then, from  Eq.~\eqref{alphaeetaBound}, 
\begin{equation}
	|\ba -1| \lesssim B_\gamma \, .
\end{equation}
 In conclusion, for systems with arbitrary $\ba$ values, it is not correct to assume that the $\gamma$ bounds should be applied to $\eta$.

As previously discussed, if $\ba$ is a constant, then it is possible to redefine $U$ (or $\rho$) to absorb $\ba$. This procedure does not leave the bound \eqref{alphaeetaBound} invariant. The reason being that one is redefining the  mass (or the gravitacional constant), and such  redefinition does change the light trajectories and the Shapiro time delay, but it does not change $\eta$,  since $\eta$ does not depend on the gravitational constant. This simple observation has the following consequence: in general, it is not possible to infer conclusions of a complete system by subdividing it into parts in which $\ba$ and $\eta$ are approximate constants. This since  each part will have its own mass definition (or its own gravitational constant). 

It is important to stress a (natural) limitation of the PPN parametrization: for systems without a proper Newtonian limit, $\ba$ cannot be set to be 1, hence neither there is a well defined  $\gamma$, in the sense of the \wn~metric,  nor one can assume that $\eta$ will be subjected to the $\gamma$ bounds. For instance, when studying external galaxies and without assuming GR, a Newtonian limit need not to exist, and one may consider relaxing the Newtonian limit. By doing so, the PPN parameter $\gamma$ is not  well defined. In the latter case, one can talk about $\gbg$, $\bg$ and $\eta$, but not $\gamma$ [since $\gamma$ is defined from the metric \eqref{ppnmetric}].\footnote{This definition for $\gamma$ is clear and not prone to generate misinterpretations.} For instance,  Ref.~\cite{Collett:2018gpf} (and similar works) claim to find $\gamma$ limits from external galaxies. In its context, GR is not assumed, but the Newtonian limit is (i.e., $\ba=1$). Together with the assumption that $\eta$ is constant inside the galaxy being studied, they indeed find bounds for $\gamma$ from the analysis of $\eta$; since with these hypothesis $\eta$ and $\gamma$ coincide.

At last, there is also an issue with nomenclature, which can lead to some confusion. Although we follow here a common convention for the slip ($\eta$) \cite{Daniel:2009kr, Amendola:2012ys, Aghanim:2018eyx, Ishak:2018his}, there are other conventions. For instance, Ref.~\cite{Amendola:2016saw} uses the same symbol $\eta$ with the same definition that we are using here in momentum space \eqref{slipk}, but it uses the name gravitational slip for $\varpi \equiv 1/\eta -1$, following \cite{Caldwell:2007cw} and similarly to \cite{Baker:2012zs}. There are also works that use the symbol ``$\gamma$'' for $\phi/\psi$  but consistently use it, without trying to impose the standard PPN bounds on it (e.g., \cite{Baker:2012zs, Baker:2011jy}).


\section{The gamma parameter in generalized Brans-Dicke theories and $f(R)$ theories}\label{sec:st}

\subsection{Generalized Brans-Dicke before considering the Newtonian limit}

In this section we consider the PN limit of generalized Brans-Dicke theories, that is, Brans-Dicke theories with a potential. These theories are among the simplest GR generalizations, but yet they are sufficiently nontrivial for a practical discussion on the gamma parameters and the gravitational slip. Furthermore, they will be useful to explore $f(R)$ theories, which are considered in Section~\ref{sec:fr}. At last, the results here shown will be generalized toward Horndeski theories in  Section \ref{sec:horndeski}.

We start by considering the following action,
\begin{equation}
S=\int\dfrac{\sqrt{-g}}{2\kappa}\left[\Phi R+ 2\dfrac{\omega{\scriptstyle (\Phi)}}{\Phi}X-V{(\Phi)}\right]d^{4}x + S_m, \label{BDaction}
\end{equation}
where $S_m$ is the action of the matter fields (assumed to be independent from $\Phi$), $\kappa$ is the coupling constant (which is here dimensionless, since we are using  $G = 1$), and 
\begin{equation}
	X=-\,\frac{1}{2}\,g^{\mu\nu}\p_{\mu}\Phi\,\p_{\nu}\Phi 
\end{equation}
is the kinetic term. There are two free functions in the theory, the scalar field potential $V$ and the coupling function $\omega$.

Variations with respect to the metric and the scalar field return the following field equations,
\begin{align}
G_{\mu\nu} &= \ \dfrac{\kappa}{\Phi}\,T_{\mu\nu} +\dfrac{1}{\Phi}\left[\nabla_{\nu}\nabla_{\mu}\Phi-g_{\mu\nu}\Box\Phi\right] \ + \notag\\[1ex]
& \quad \ + \dfrac{\omega}{\Phi^2}\left[\p_\mu\Phi\p_\nu\Phi+Xg_{\mu\nu} \right] -g_{\mu\nu}\,\dfrac{V}{2\Phi},\label{metriceq}\\[2ex]
\dfrac{2\omega}{\Phi}\,\Box\Phi &=-R + \dfrac{2\omega'}{\Phi}X -\frac{2\omega}{\Phi^2}X +V',\label{phieq}
\end{align}
where $G_{\mu\nu}$ is the Einstein tensor, $T_{\mu\nu}$ is the usual energy-momentum tensor, $\nabla_{\mu}$ indicates a covariant derivative, $\Box=\nabla_{\mu}\nabla^{\mu}$ is the d'Alembertian operator and the prime symbol $``~'~"$ represents a derivative with respect to the scalar field $\Phi$.

A consistent PN expansion up to $O(2)$ is obtained by considering\footnote{The relation $A \sim O(n)$  means that $|A|$ is at most of the same order of $v^n$, that is $|A| \lesssim v^n$.}
\begin{equation}
\Phi = \varphi_{0}+\varphi,\quad\mbox{with } \varphi_0 > 0  \mbox{ and } \varphi \sim O(2) . \label{Phiphi}
\end{equation}
The $\varphi_0$ term is the zeroth order expansion, i.e. $\varphi_0 \sim O(0)$, and  it is the average value of $\Phi$ in the considered spacetime region in the absence of the local system being considered (i.e., $\varphi_0$ depends on other systems that are much larger than the system being considered, cosmology for instance).\footnote{Consequently, $\varphi_0$  does not depend on either $t$ or $\mathbf x$, but depends on the central time and space values selected to do the expansion.}  Therefore, within the considered system, $\Phi$ must asymptotically approach $\varphi_0$ as one moves farther from its center. The relevant length scale of the system  should be small enough such that the true metric can be written as an expantion about Minkowsk, as in Eq.~\eqref{barmetric}. The value of $\varphi_0$ is taken as positive to guarantee the Newtonian limit.

Henceforth  we consider that $V(\Phi)$ and $\omega(\Phi)$ are either analytical functions of $\Phi$ or that they can be approximated by 
\begin{align}
&V(\Phi)\a V_0 + V_1\varphi +V_2\varphi^{2}, \label{Vphiexpand}\\
&\omega(\Phi) \approx  \omega_0 + \omega_1\varphi +\omega_2\varphi^{2},\
\end{align}
with $V_n$ and $\omega_n$ constants. In particular, $V(\varphi_0) = V_0$  and $\omega(\varphi_0) = \omega_0$, since $\Phi=\varphi_0$ implies $\varphi=0$.

From Eq.~\eqref{metriceq}, one sees that $V_0$ acts as an effective cosmological constant. Since far from the Sun the PPN metric must approach the Minkowski metric, $V_0$ needs to be sufficiently small such that it can be neglected in the solar system.  It is not possible to consider $V_0$ as non-negligible in the WN-PPN formalism without a careful and specific justification, this  since it violates one of its cornerstones assumptions: asymptotic flatness \cite{Will:1993ns, Will:2014kxa}. The effect of a cosmological constant to the bending of light requires a careful and specialised analysis which is not covered  by this formalism (e.g., see the discussion in refs.~\cite{Rindler:2007zz,Biressa:2011vy,Piattella:2015xga,Faraoni:2016wae}).

The matter content is, up to the considered order, approximated by a perfect fluid description, with
\begin{align}
	T^{\mu\nu}=\left(\rho+\rho\Pi+p\right)u^{\mu}u^{\nu}+pg^{\mu\nu}\,,
\end{align}
where $\rho$ is the mass density, $\Pi$ is the fluid's internal energy per unity mass, $p$ is the pressure and $(u^{\mu}) =(dx^{\mu}/d\lambda) = (u^0, u^i)=u^{0}(1, v^i)$ is the fluid four-velocity.

 Following the slow motion condition, the energy momentum tensor and metric components are expanded in orders of $v \sim O(1)$  and one finds \cite{Will:2014kxa} 
\begin{equation}
\rho\sim O({2})\quad\mbox{and}\quad p\sim \rho\Pi\sim O({4}).
\end{equation}

It is also well known that  time derivatives  increase the perturbation order by one (i.e. $\p_t\sim O(1)$); and that, from the viral theorem applied to Newtonian gravity, it is known that the Newtonian gravitational potential is of order $O(2)$, which must be the largest contribution to the metric perturbations, denoted by $h_{\mu \nu}$. Hence,
\begin{equation}
h_{\mu\nu}\sim O({2}). 
\end{equation}

To solve the modified Einstein's equations \eqref{metriceq}, the Ricci tensor is expanded up to the second order, $O(2)$, as follows [using Eq.~\eqref{barmetric}], 
\begin{align}
	R_{00}&\a - \nabla^{2}(\ba \,U),\\[1ex]
	R_{0i}&\a 0\\[1ex]
	R_{ij}&\a -\delta_{ij}\nabla^{2}(\bg \,U) +  \partial_{j}\partial_{i}[(\ba +\bg)U].
\end{align}
We recall that $U$, $\ba$ and $\bg$  depend on time in general, but their time derivatives do not appear in the expressions above since they are of order $O(3)$ or higher order.

The energy momentum tensor, up to second order, has a single non-null component given by  $T_{00}\a \rho$. Hence, using the field equations \eqref{metriceq}, the  Ricci scalar can be written as 
\begin{equation}\label{ricci}
	R\a \frac{\kappa}{\varphi_0}\,\rho +\frac{3\nabla^2\varphi}{\varphi_0}+\frac{2V_1}{\varphi_0}\,\varphi. 
\end{equation}

Therefore, Eq.~\eqref{phieq} up to $O(2)$ reads
\begin{align}\label{helmholtz}
\nabla^{2}\varphi - m_{\varphi}^{2}\varphi= -\dfrac{\kappa\rho}{3+2\omega_0},
\end{align}
with the mass term given by,
\begin{equation}\label{mass}
m_{\varphi}^2=\dfrac{2\left(V_2\varphi_0-V_1\right)}{3+2\omega_0}.
\end{equation}
This mass expression can also be found in Refs.~\cite{Olmo:2005hc, Capone:2009xk}. The special case  $\omega_0=-3/2$ will be considered in Section \ref{sec:palatini}.

Equation \eqref{helmholtz} solution, with the boundary condition that $\varphi$ should approach zero far from the Sun, is a Yukawa potential, which is expressed as follows,
\begin{equation}\label{yukawa_st2}
\varphi= \frac{\kappa}{4 \pi (3 + 2 \omega_0)}\int\frac{\rho(x',t)}{\x}\,e^{-m_{\varphi}\x}\,d^{3}x'.
\end{equation}

Signatures of an Yukawa correction to the Newtonian potential have been examined by laboratory experiments and astronomical observational tests. Up to now, no deviations from Newton's inverse square law have been detected from sub-millimeter to astronomical distances (see \cite{Kapner:2006si,Newman:2009zz,Adelberger:2003zx, Westphal:2020okx} and the references therein).

Using Eqs.~(\ref{u},  \ref{ricci}, \ref{helmholtz}), the time-time component of Eq.~\eqref{metriceq}  up to $O(2)$ yields 
\begin{equation}
	\frac{16\pi\varphi_0}{\kappa}\,\nabla^2(\ba U)= \ a_1 \,\nabla^2U   +a_2 \nabla^2\varphi \, , \label{alphaUaphi}
\end{equation}
where 
\begin{align}
a_1 \equiv 2 - \frac{2}{3 + 2 \omega_0} \frac{V_1}{m_\varphi^2}
\mbox{ and }
a_2 \equiv \ \frac{8 \pi}{\kappa} \left(1+\frac{V_1}{m^2_\varphi}\right).
\end{align}

Therefore, Eq.~\eqref{alphaUaphi} can be solved as
\begin{align}
	\ba=& \ \frac{\kappa}{16\pi \varphi_0}\left(a_1+ a_2\,\frac{\varphi}{U}\right)\, .\label{h00}
\end{align}
The relation above must be satisfied always, independently on the Newtonian limit.

In order to find $\gamma$ and $\eta$, we also need the field equations for the spatial components. Writing  $g_{ij} = \delta_{ij} + h_{ij}$, up to first order on $ h_{ij}$,  we consider the gauge
\begin{equation}\label{gauge-hij}
		\p_kh^k{}_{i}+\frac{1}{2}\,\p_ih_{00}-\frac{1}{2}\,\p_ih^k{}_k=\frac{1}{\varphi_0}\,\p_i\varphi. 
\end{equation}
Therefore, the spatial part of Eq.~\eqref{metriceq} can be written as
\begin{equation}
\frac{16\pi\varphi_0}{\kappa}\,\nabla^2(\bg U)=  a_3 \nabla^2U  - a_2 \nabla^2\varphi \, ,
\end{equation}
with
\begin{equation}
	a_3 \equiv 2 +  \frac{2}{3 + 2 \omega_0} \frac{V_1}{m_\varphi^2} = 4 - a_1 \, .
\end{equation}

From the above it is found that
\begin{align}
\bg= \ \frac{\kappa}{16\pi \varphi_0}\left(a_3- a_2\frac{\varphi}{U}\right).\label{hij}
\end{align}

The standard PPN bounds cannot be applied to $\bg$, as previously explained. However, one can compute $\gbg$ from Eq.~\eqref{gammaBG}, which is a constant for this case, namely
\begin{equation} 
	\gbg =\frac{\kappa}{4\pi \varphi_{0}} -1. \label{gamma-sc}
\end{equation}

The (physical space) slip parameter is computed from  Eq.~\eqref{slipdeff},
\begin{equation}
	\eta =\dfrac{a_3-a_2\varphi/U}{a_1 +a_2\varphi/U}  = - 1 + \frac{4}{a_1 +a_2\varphi/U} \, .\label{slip-st}
\end{equation}
For a fixed value of  $\kappa$, in the limit $\omega_0 \rightarrow \infty$ or $m_\varphi \rightarrow \infty$, one finds $\eta \rightarrow 1$, as expected.\footnote{The limit $\omega_0 \rightarrow \infty$ implies that $m_\varphi \rightarrow 0$, however $m_\varphi \rightarrow 0$ does not imply that $\omega_0 \rightarrow \infty$. We can parametrize the model with $m_\varphi$ as an independent parameter, in place of $V_2$.}

The above equations stress  the differences between $\bg, \; \gbg$ and $\eta$. As explained in Secs.~\ref{sec:gbg} and \ref{sec:slipIsNotGamma}, the standard PPN bound on $\gamma$ can be used on $\gbg$, not on $\eta$.  The $\gbg$ parameter in generalized Brans-Dicke theories is always a constant and its value depends on the theory's coupling constant $\kappa$. In particular this means that its numerical value  is influenced by  Newtonian limit, or, in the absence of a Newtonian limit, by the definition of the mass (gravitational constant). The gravitational slip $\eta$, on other hand, is a spacetime function whose value is independent from $\kappa$.

In the following three subsections, we consider specific developments  according to the mass of the scalar field. Metric and Palatini $f(R)$ theories are considered in subsections \ref{sec:fr} and \ref{sec:palatini}.  We comment on other results in the literature about $\gamma$ and $\eta$ in subsection \ref{sec:BDdiscussion}.

\subsection{Scalar field with negligible mass} \label{sec:BDnegligible}
Here we are concerned with the case in which $m_\varphi$ is small enough compared with the system length scale. We start by expanding the exponential in Eq.~\eqref{yukawa_st2} about $m_\varphi =0$. Hence,
\begin{equation}
	\varphi \propto \int\frac{\rho(x',t)}{\x}\, [1  -m_{\varphi}\x + O(m^2_\varphi)]\,d^{3}x' \, .
\end{equation}
The first term of the expansion above states that $\varphi \propto U$, and it leads to a well posed Newtonian limit. The second term does not breaks the previous result, since it simply adds a constant to $\varphi$ (or $U$), which will not change the Newtonian picture. The third term adds a function that has no counterpart in Newtonian physics. Hence, here we consider $m_\varphi$ to be small enough such that the third term is negligible; otherwise there would be no Newtonian limit in the considered system. For the order expansion used in PPN, and if $\ell$ represents the typical length of the system, we have $m_\varphi^2 \ell^2 \sim O(1)$.

Consequently, for a given mass $m_\varphi$, and inside the system of length scale $\ell$, with $m^2_\varphi \ell^2 \sim O(1)$, $\ba$ behaves approximately as a constant, namely
\begin{equation}\label{h00massless}
\ba\approx\frac{\kappa}{4\pi \varphi_{0}} \frac{2+\omega_0}{3+2\omega_0} \, .
\end{equation}
The approximation symbol above is used to emphasize that Eq. \eqref{h00massless} does not hold in general, since for sufficiently large distances $\ba$ is not a constant. We already fixed $G =1$ [see Eq.~\eqref{u}], hence it is not possible to consider a $G$ redefinition in order to absorb the $\ba$ value above. Therefore, the Newtonian limit demands the right hand side of \eqref{h00massless} to be equal to $1$, which implies 
\begin{equation}\label{newt-limit-massless}
\frac{\kappa}{4\pi\varphi_{0}}=\frac{3+2\omega_0}{2+\omega_0}\, .
\end{equation}
The relation above, combined with Eq. \eqref{gamma-sc}, results in
\begin{equation}
	\gbg=\frac{1+\omega_0}{2+\omega_0}.\label{gama_massless}
\end{equation}
It is worth to reinforce that expression \eqref{gama_massless} is valid even where the Newtonian limit is not valid.

Within the approximation above $\bg$ will also behave approximately as a constant inside the considered system, and the EPPN metric \eqref{barmetric} achieves the same form of the \wn~one \eqref{ppnmetric}. \textit{In this limit}, all the $\gamma$'s and the slip are the same, since
\begin{equation}
	\eta=\bg=\gamma=\gbg	 \, \mbox{ \small (for the \wn~metric).}
\end{equation}
On the other hand, in general, $\bg$, $\gbg$ and $\eta$ can be used and computed independently from the Newtonian limit, and their meanings and bounds are in general independent from the $\gamma$ bounds, as shown in Sections~\ref{sec:gbg} and \ref{sec:slip}.

Since the expressions used above include inverse powers of $m_\varphi$, we cannot consider the case $m_\varphi=0$ exactly. However, one can directly verify that the cases  $m_\varphi \to  0$ or $m_\varphi=0$ lead to the same results for $\gbg$. The case $m_\varphi=0$ was considered in several works, see e.g. \cite{Will:1993ns}.
The distinction is reserved to the fact that, for the massless case ($m_\varphi=0$), all the gamma’s and the slip are constants at any point in space, while for the approximate case ($m_\varphi\approx 0$) in general there will be sufficiently large distances such that these parameters will change, and become different among themselves.

\subsection{Large mass scalar field} \label{sec:BDlargemass}

From Eq.~\eqref{h00}, we can write
\begin{equation}
	g_{00} = -1 + \frac{\kappa}{8 \pi \varphi_0}(a_1 U + a_2 \varphi) \, . \label{g00BD}
\end{equation}
The three options for finding a Newtonian limit are that either $\varphi$ is proportional to $U$, $\varphi$ is a constant or $\varphi$ is negligible, that is $| \varphi|  \ll U$. From Eq.~\eqref{yukawa_st2}, one sees that the latter can be found if $m_\varphi$ is sufficiently large (i.e., $m_\varphi^2 \ell^2 \gg 1$). More precisely, we need that $\varphi \sim O(3)$ or $e^{- m_\varphi \ell}\sim O(1)$ (PN considerations will enforce this limit to be stronger, but from the Newtonian physics alone that is sufficient). Assuming this case, the term $a_2 \varphi$ in the above equation becomes negligible within the considered system and, from the Newtonian limit, one must impose the relation
\begin{equation}
	\frac{\kappa}{4 \pi \varphi_0} =\frac{4}{a_1}\, .\label{newt-limit-massive}
\end{equation}

Hence,  $\gbg$ can be computed from Eq.~\eqref{gamma-sc}, implying that
\begin{equation}
	\gbg=\frac{4}{a_1}-1= \frac{a_3}{a_1}= \frac{(3 + 2 \omega_0) m_\varphi^2 + V_1}{(3 + 2 \omega_0) m_\varphi^2 - V_1} \, . \label{gbgBDmassive}
\end{equation}

Likewise the previous case with negligible mass, we have succeeded in re-writing the metric in the same form of the standard PPN metric, hence all the gamma's and the slip are the same in the considered regime ($e^{-m_\varphi \ell} \sim O(1)$). On the other hand, for distances much smaller than $\ell$, the Newtonian correspondence is broken and $\eta$ and $\bg$ deviates from $\gbg$.

The GR result $\gbg = 1$ is found if either $m_\varphi \rightarrow \infty$, $\omega_0 \rightarrow \infty$ or $V_1 \rightarrow 0$. The case $V_1=0$ is commonly studied and it yields $\gbg=1$ for the large mass case. The limit $m_\varphi \rightarrow 0$ cannot be used in the above, since we are considering $e^{- m_\varphi \ell} \sim O(1)$.

\subsection{Intermediate mass scalar field} \label{sec:BDintermediate}

It is tempting to ask what happens if the scalar field mass is neither negligible nor large. More precisely, for a system that is evaluated within a typical length scale $\ell$, if $m_\varphi$ satisfies 
\begin{equation}
	 O(1/2) < m_\varphi \ell 	< |\ln O(1)| \, 
\end{equation}
what would be the consequences? From the two previous sections, one sees that, for such case, $\varphi$ is neither proportional to $U$ plus a constant, nor it is negligible, hence $h_{00} \not \propto U + \mbox{constant} + O(3) $. Consequently, there is no well defined Newtonian limit for the considered system and  one cannot truly apply a PPN formalism (as the formalism name suggests).  We recall that a principle of the PPN formalism is that any non-Newtonian correction should appear \textit{beyond} the Newtonian order, never at the Newtonian order. 

The issue above is  developed in detail by Alsing \textit{et al} \cite{Alsing:2011er}, where the observational data from the Shapiro time delay, due to Cassini, are used to constrain the  massive Brans-Dicke theory. Hence, constraints on the allowed regions for the pair ($\omega_0 , m_\varphi$), in the intermediate mass range, can be estimated considering the selected observational data. As developed in that reference (in their Section IV), it is not possible to fully follow the \wn \, formalism in the intermediate mass range. They used a modified PPN formalism in which $U$ includes a Yukawa correction. It is shown that the Shapiro time delay is exactly that of GR, apart from a constant multiplicative factor that is related to the mass definition of the Sun (an issue that deserves special care in theories without a Newtonian limit). It is curious that, following their conventions, this constant factor is written as an effective gamma parameter denoted by $\tilde \gamma$, which has the same form of the slip,\footnote{We note that this reference uses $V_1 =0$, while we left $V_1$ free in order to compare with other references that do not fix it.} but with the Earth's  orbit radius in place of the radial $r$ dependence (due to how the mass is defined in their approach, not due to an approximation within given radial range). That is, contrary to the slip, $\tilde \gamma$ needs to be a true constant. 

The result of Ref.~\cite{Alsing:2011er} is compatible with ours, which is based on $\gbg$ \eqref{gamma-sc}. They have found that the the Shapiro time delay behaves exactly as in GR, apart from a constant redefinition of the mass of the Sun. This is precisely what a constant $\gbg$ means (and we recall that $\gbg$ does not depend on the Newtonian limit to be a valid parametrization for light deflection and the Shapiro time delay). Moreover, following their conventions we can reproduce their results, which for completeness we do below.

From eq.~\eqref{g00BD} with $V_1 =0$, $\rho(\mathbf x) = M_\odot \delta^{3}(\mathbf x)$ and $|\mathbf x| = r$, we find 
\begin{align}
	h_{00} & = \frac{\kappa  }{4 \pi \varphi_0} U  + \frac{1}{\varphi_0} \varphi  \nonumber \\  
	& = \frac{\kappa}{4\pi \varphi_0} \frac{M_\odot}{r} \left( 1 + \frac{1}{3+ 2 \omega_0} e^{-m_\varphi r}\right) \, .
\end{align}

Recalling that we have already used $G=1$, and following the principles of Ref.~\cite{Alsing:2011er} (see also  Ref.~\cite{Perivolaropoulos:2009ak}), now we set the coupling constants such that Earth's orbit follows Kepler laws (this approach simply focus on the Shapiro time delay, and neglects other observables, like the precise orbits of the other planets). Anyway, this can be assured if 
\begin{equation}
	h_{00}(r_{\s \oplus}) = 2 \frac{M_\odot}{r_{\s \oplus}}, 
\end{equation}
where $r_{\s \oplus}$ stands for the Earth's orbit radius. Therefore,
\begin{equation}
	\frac{\kappa}{4 \pi \varphi_0} = \frac{ 2} {1+ \frac{1}{3 + 2 \omega_0} e^{- m_\varphi r_{\s \oplus}}}\, 
\end{equation}
and, from eq.~\eqref{gamma-sc},
\begin{equation}
	\gbg = \frac{1- \frac{1}{3 + 2 \omega_0} e^{- m_\varphi r_{\s \oplus}}} {1+ \frac{1}{3 + 2 \omega_0} e^{- m_\varphi r_{\s \oplus}}} \,  = \tilde \gamma \, .
\end{equation}
In the above, $\tilde \gamma$  is the same used in Ref.~\cite{Alsing:2011er}. One can also state that $\eta(r_{\s \oplus}) = \tilde \gamma$.

Reference \cite{Alsing:2011er} was not the first to study the intermediate mass case,  see in particular Refs.~\cite{Perivolaropoulos:2009ak, Capone:2009xk}. These two  cite  the approach of Refs.~\cite{Olmo:2005hc, Olmo:2005zr} for matching an effective gamma with the slip. Perivolaropoulos \cite{Perivolaropoulos:2009ak} adopted a convention that lead, under certain approximation, to the same bounds found in Ref.~\cite{Alsing:2011er}. Indeed, both works show equivalent allowed regions in the $(\omega, m_\varphi)$ plane.\footnote{The plots in these references are compatible, but they  consider different independent variables: while Ref.~\cite{Alsing:2011er} uses $m_s$ (which is ours $m_\varphi$), Ref.~\cite{Perivolaropoulos:2009ak} uses $m_0 \propto m_\varphi \sqrt{2\omega + 3 }$.} While Ref.~\cite{Perivolaropoulos:2009ak} uses $r \approx 1$ au as an approximation valid for any solar system observation, Ref.~\cite{Alsing:2011er} uses $r = 1$ au  since ``it is the scale associated with the determination of the Keplerian mass of the Sun''. Hence, for the latter, it is a matter of defining units, the effective gamma ($\tilde \gamma$ or $\gbg$) is not approximately a constant, it is a true constant, no matter what are the conventions used to define the Keplerian gravitational mass.  If one simply uses $\eta(r)$ in place of $\gamma$, one would find that the Shapiro time delay would acquire a dependence on the impact parameter different from that of GR,\footnote{Namely, the Shapiro time delay, either in GR or in generalized Brans-Dick theories, decays with $\ln(b^{-2})$, where $b$ is the impact parameter \cite{Alsing:2011er}.} which is false.

The points  to be stressed are that: i) there is a constant parameter $\gbg$ that describes light trajectories and time delay, just like the standard $\gamma$, which is different from $\eta(r)$. ii) Physical constraints within the intermediate mass regime can be found, but they are not part of a standard PPN approach, since they violate the Newtonian limit.

\subsection{Metric $f(R)$ theories}\label{sec:fr}
The extension of GR by including non-linear corrections of the Ricci scalar in the action is a well known proposal for modified gravity theories (see e.g., \cite{Sotiriou:2008rp, Capozziello:2010zz, Nojiri:2017ncd} for reviews on the subject). Since there is an equivalence between $f(R)$ and scalar-tensor theories, the analysis of the previous subsections is tightly related to such theories, both in metric and Palatini formulations \cite{Sotiriou:2008rp}. 

This class of theories is defined from the action
\begin{equation}
	S = \frac{1 }{2 \kappa}\int f(R) \sqrt{-g}\, d^4x + S_m \, .
\end{equation}
In order for the PN expansion on $v$ to be consistent and with a viable Newtonian limit, similarly to $V(\Phi)$ in scalar-tensor theories, $f(R)$ cannot be an arbitrary function.  In the considered system, it should be possible to expand it as follows \cite{Sotiriou:2008rp, Capozziello:2010zz} 
\begin{equation}
	f(R) = f_0 + f_1 R + f_2 R^2 + O(R^3) \, ,	\label{fRexpand}
\end{equation}
where $f_i$ are  constant coefficients. The higher order corrections $ O(R^3)$ are not relevant for the developments here presented. Since PPN requires asymptotic flatness, we henceforth consider that $f_0$ is sufficiently small to be negligible for the dynamics.

The interpretation of a $f(R)$ model as  an effective scalar-tensor theory is done through the following Legendre transform,
\begin{align}
	& \Phi \equiv \frac{df(R)}{dR},\label{frphi}\\[1ex]
	& V(\Phi) \equiv R \Phi - f(R)  \, . \label{frVphi}
\end{align}

Whenever the Hessian $d^2f/dR^2$ is non singular, this transformation is both well posed  and it has an inverse. Therefore, all the information in the scalar-tensor frame should be equivalent (within a given map) to the $f(R)$ one. For more complex cases with several scalars and singular Hessians, see Ref.~\cite{Rodrigues:2011zi}. 

It is well known that the resulting action is a particular case of the scalar-tensor action \eqref{BDaction}, with $\omega = 0$. Moreover, from Eqs.~(\ref{Phiphi}, \ref{fRexpand}, \ref{frphi}) one finds
\begin{equation}\label{phi-fr}
	\varphi_0 = f_1 \; \mbox{ and } \; \varphi = 2 f_2 R \, .
\end{equation}
And from these results together with Eqs.~(\ref{Vphiexpand}, \ref{mass}),
\begin{equation}\label{V-fr}
	V_1 = 0, \;\; V_2 = \frac{1}{4f_2} \; \mbox{ and } \; m^2_\varphi= \frac{1}{6}\frac{f_1}{f_2}\,.
\end{equation}
Had we kept an arbitrary $f_0$, at this point we would find $V_0 = f_0$, but since $V_0 =0$, we would conclude $f_0=0$.

 Just like the scalar-tensor case, a Newtonian limit can only exist if either $f_1 \ell^2 /(6f_2)  \sim O(1)$ or $e^{- \sqrt{f_1 \ell^2 / (6 f_2)} } \sim O(1)$, which, for fixed $f_1/f_2$, depends on the typical size ($\ell$) of the system under consideration. In a PN context, the Newtonian limit should be imposed for one of these cases and, consequently, the possible answers for $\gbg$ in $f(R)$ theories are,
  \begin{equation}
 	\gbg = 
 	\begin{cases}
 			\frac{\kappa}{4\pi f_1}-1\,, \; \mbox{in general}\, ;\\[.2cm]
 		\frac{1}{2} \, , \; \mbox{ if } \frac{f_1 \ell^2}{6f_2}  \sim O(1)\, ;  \\[.2cm]
 		1 \;,  \; \mbox{ if }	e^{- \sqrt{f_1 \ell^2 / (6 f_2)} } \sim O(1) \, .
 	\end{cases}
 \end{equation}
 These values are obtained by imposing $f(R)$-equivalent conditions \eqref{newt-limit-massless} and \eqref{newt-limit-massive} and using Eq. \eqref{gamma-sc}.
 In the regions where there is a Newtonian correspondence, the metric assumes the PPN form and one can obtain the equivalent $\gamma$, otherwise it is not defined. Thus, only two possible answers can emerge:
\begin{equation}
	\gamma = 
	\begin{cases}
		\frac{1}{2} \, , \; \mbox{ if } \frac{f_1 \ell^2}{6f_2}  \sim O(1)\, ;  \\[.2cm]
		1 \;,  \; \mbox{ if }	e^{- \sqrt{f_1 \ell^2 / (6 f_2)} } \sim O(1) \, .
	\end{cases}
\end{equation}

The slip comes from Eq.~\eqref{slip-st} and it reads
\begin{equation}
	\eta = \frac{1 - \frac{1}{3} \xi}{1 + \frac{1}{3} \xi} \, ,
\end{equation}
with
\begin{equation}
	\xi \equiv \frac{\int\frac{\rho(x',t)}{\x}\,e^{-\sqrt{\frac{f_1}{6 f_2}}\x}\,d^{3}x'}{\int\frac{\rho(x'',t)}{\xx}\,d^{3}x''} \, .
\end{equation}
If all the mass is in a single particle (p) at $\mathbf x_0$, it is simply 
\begin{equation}
	\xi|_{\rm \s p} = e^{-\sqrt{\frac{f_1}{6 f_2}} |\mathbf x - \mathbf x_0| }\, .
\end{equation} 
Although the true gamma ($\gamma$) can only have the values 1 or 1/2 for $f(R)$ theories, $\eta$ is not a constant. The approximately constant cases $\eta \approx 1/2$ and $\eta\approx1$ occur at distances much smaller and much larger than $\ell$, respectively. More precisely, one has
 \begin{equation}
	\eta = 
	\begin{cases}
	\mbox{\small Spatial function without direct}\\ \mbox{\small relation to light deflection, in general}\, ; \\[.2cm] 
		\frac{1}{2} \, , \; \mbox{ if } \frac{f_1 \ell^2}{6f_2}  \sim O(1)\, ;  \\[.2cm]
		1 \;,  \; \mbox{ if }	e^{- \sqrt{f_1 \ell^2 / (6 f_2)} } \sim O(1) \, .
	\end{cases}
\end{equation}
The slip can be very important to describe modified gravity, but it does not have the same role of $\gbg$ or $\gamma$. The slip is a comparison between the two potentials $\phi$ and $\psi$; being such that GR satisfies $\eta=1$. On the other hand, both $\gamma$ and $\gbg$ parametrize electromagnetic waves. In the regions where there is a Newtonian limit, $\eta$, $\gamma$ and $\gbg$  coincide. In general, however, $\eta$ does not parametrize light deflection, nor the gamma bounds have any meaning for $\eta$.
 
\subsection{Palatini $f(R)$ theories} \label{sec:palatini}

Palatini formulations are a special case where $\omega=-3/2$, implying that the scalar field is not dynamical. One must return to Eq. \eqref{helmholtz} and rewrite it to obtain that the scalar field is now proportional to the fluid mass density, namely
\begin{equation}
\varphi=\frac{2\kappa f_2}{f_1}\,\rho,
\end{equation}
after using expressions \eqref{phi-fr} and \eqref{V-fr}.
Apart from the above distinction, the results for $\ba$ and $\bg$ continue to hold since the $\omega$-proportional terms in the field equations \eqref{metriceq} have no influence up to $O(2)$. One just needs to eliminated the mass terms, through Eq.~\eqref{mass}, in the $a_n$ constants before substituting $\omega_0=-3/2$, which gives [together with \eqref{V-fr}],
\begin{align}
&a_1=a_3=2,\\[1ex]
&a_2|_{\s \omega_0=-\frac{3}{2}}=\frac{8\pi }{\kappa}.
\end{align}
Using the above  and Eq. \eqref{h00}, 
\begin{equation}
	g_{00}=-1+\frac{\kappa}{4\pi f_1}\,U + \frac{2\kappa f_2}{f_1^2}\,\rho.
\end{equation}
The last term indicates that Newtonian gravity is violated inside matter. However, this violation is relevant only if the pressure and the internal energy are known from first principles (not indirectly through gravitational effects), since in principle it is possible to redefine these quantities and completely absorb the non-Newtonian contribution above (as we have shown in detail in Ref.~\cite{Toniato:2019rrd}). Thus, apart from the case with microscopic modeling for the fluid pressure and internal energy, which is not a common situation for the systems considered under PPN,  the Newtonian limit of Palatini $f(R)$ theories is well posed if one sets
\begin{equation}\label{f1}
	f_1=\frac{\kappa}{8\pi}.
\end{equation}
Consequently, it is obtained
\begin{equation}
	\gbg=1,
\end{equation}
and light trajectories are the same as in GR. It is easy to see that, in vacuum, the linearized metric assumes the usual PPN form (since $f(R)$ Palatini is identical to GR in vacuum) and, with Eq. \eqref{f1}, one also finds $\gamma=1$.

\subsection{Discussion on the $f(R)$ and Brans-Dicke results} \label{sec:BDdiscussion}

In the context of scalar-tensor and $f(R)$ gravity, several references do a nice work on the computation of the slip, which is relevant for several bounds and analyses, but use the slip (or suggest that it can be used) as if it had the same physical properties of $\gamma$ and, therefore,  as if it would be submitted to be same numerical constraints. It is not our purpose to present an exhaustive list of such issues. A few cases include Refs.~\cite{Olmo:2005hc, Olmo:2005zr, Kainulainen:2007bt,  Perivolaropoulos:2009ak, Capone:2009xk, Hohmann:2013rba,  Sbisa:2018rem, Faraoni:2019sxw}. 
 
References \cite{Olmo:2005hc, Olmo:2005zr, Hohmann:2013rba, Faraoni:2019sxw}  compute the slip and use in place of $\gamma$ (sometimes called effective $\gamma$), without providing further details. In part, they are simply computing the slip with a notation different from ours, which is valid and relevant in their contexts; but it is also suggested that the gamma they compute is the one from PPN, which would be subjected to the standard $\gamma$ bounds. References \cite{Perivolaropoulos:2009ak, Capone:2009xk} cite \cite{Olmo:2005hc, Olmo:2005zr} in part to justify their procedures; while Ref.~\cite{Sbisa:2018rem} justify their use of $\eta$ in place of $\gamma$ in part by  citing \cite{Hohmann:2013rba}. All such procedures may perhaps be justified under certain limits,  together  with precise considerations on the gravitational mass definition whenever there is no explicit Newtonian limit. 

The case of Ref.~\cite{Perivolaropoulos:2009ak} is particularly interesting, since the derived numerical bounds were correct, but a detailed explanation on why it really works (and the limitations of such specific approach) only appeared latter \cite{Alsing:2011er} (Section \ref{sec:BDintermediate} details further this point). In essence, the results of Ref.~\cite{Perivolaropoulos:2009ak} are correct if the slip is not assumed to be an approximate constant equal to $\gamma$, but if instead one first properly define the gravitational mass, leading to a (truly) constant effective gamma which  indeed parameterizes both light deflection and the time delay (this constant coincides with $\gbg$, as here used). This approach, with constant $\gbg$,  has important consequences that will be further commented in the next sections.

Reference \cite{Sbisa:2018rem} considers pressure effects to the PPN $\gamma$ parameter. In a proper \wn , approach it is not possible for the pressure to have an effect on $\gamma$. The reason being that pressure enters in the PPN formalism as a $O(4)$ correction, hence neither it has an impact on the Newtonian limit nor it can directly affect  $\gamma$. On the other hand, pressure can affect the slip, as the authors have shown\footnote{More precisely, what they call $\gamma$, following \cite{Hohmann:2013rba}, is a kind of slip  defined Schwarzschild coordinates.} and compared their findings to other references that also identify the slip with $\gamma$. The slip by itself is useful for several considerations and physical bounds, but it only can be compared to $\gamma$ in very special cases, since in general it does not parametrize the dynamics of light. A slip whose value is far from 1 does not imply that the dynamics of light deviates from that of GR.

This misidentification between gammas and the slip is also found in a slightly distinct context of scalar-tensor theories using different frames \cite{Scharer:2014kya}, scalar-torsion formulations \cite{Emtsova:2019qsl}, and $f(R)$ theories using an alternative method of linearization \cite{Negrelli:2020yps}.

Considering $f(R)$ gravity, specifically, we stress that it cannot have arbitrary values of $\gamma$ if a Newtonian limit is considered: it is either 1 or 1/2. This result is in agreement with Refs.~\cite{Berry:2011pb, Clifton:2008jq, Toniato:2019rrd}, in which PN analysis is considered without resorting to the scalar-tensor representation. Berry and Gair \cite{Berry:2011pb} conclude in their section VIII-A, that "$f(R)$-gravity is indistinguishable from GR" with respect to $\gamma$ parameter. They are considering models with $\kappa=8\pi$ and $f_1=1$, which recover GR when linearized with respect to Ricci scalar, and returns $\gbg=1$. However, following the discussion in Section \ref{sec:BDintermediate}, metric $f(R)$ theories violate Newtonian physics in general, and care should be taken when applying to $\gbg$ the observational bounds of $\gamma$, since the mass definition of the Sun must be corrected.

The work of Collett \textit{et al} \cite{Collett:2018gpf} is a reference work on constraining GR deviations in the context of galaxies. From the elliptical galaxy ESO 325-G004 internal dynamics and the bending of the background light, it was found that  $\phi/\psi \sim 0.97\pm 0.09$. For deriving this, it was assumed: $i$) that this ratio is approximately constant in the galaxy; and $ii$) that the internal dynamics of massive bodies come from Newtonian gravity (with a dark matter halo and a black hole). In the context of the present work, one can promptly say that this result is a constraint on the slip $\eta$. But, due to their hypothesis, it is also correct  to call it a constraint on  $\gamma$.  Considering  $f(R)$ gravity, the result from Ref.~\cite{Collett:2018gpf} strongly disfavors $\gamma = 1/2$ for that galaxy. Such results for one galaxy (likewise for the solar system) cannot be immediately extended to all galaxies, in particular a dependence on the cosmological time is in principle possible \cite{Sanghai:2016tbi, Clifton:2018cef}.

Physically, the constancy of $\gbg$ in $f(R)$ and generalized Brans-Dicke theories implies that, for a given mass, the spatial dependence of the bending of light and the Shapiro time delay are exactly the same of GR, but they may diverge on the global multiplicative factor. This observation is further generalized in the following section.

\section{Extension to Horndeski theories}\label{sec:horndeski}
The results discussed before, for generalized Brans-Dicke theories, are directly extended to the more general formulation of scalar-tensor theories according to Horndeski \cite{Horndeski:1974wa}. 

The Horndeski action is given by,
\begin{equation}\label{action}
	S=\sum_{i=2}^{5}\frac{1}{2\kappa}\int d^4x\sqrt{-g}\,{\cal L}_i + S_m,
\end{equation}
where $S_m$ is the matter action and,
\begin{align*}
	&{\cal L}_2=K(\Phi,X),\quad {\cal L}_3=-G_3(\Phi,X)\Box\Phi,\\[1ex]
	&{\cal L}_4= G_4(\Phi,X)R + G_{4X}(\Phi,X)\left[(\Box\Phi)^2- (\nabla_{\mu}\nabla_{\nu}\Phi)^2\right],\\[1ex]
	&{\cal L}_5=G_5(\Phi,X)G^{\mu\nu}\nabla_{\mu}\nabla_{\nu}\Phi -\frac{1}{6}G_{5X}(\Phi,X)\big[(\Box\Phi)^3\\
	&\qquad \ - 3\Box\Phi(\nabla_{\mu}\nabla_{\nu}\Phi)^2+2(\nabla_{\mu}\nabla_{\nu}\Phi)^3\big].
\end{align*}
In the above expressions, we have used the following notation,
\begin{align}
	(\nabla_{\mu}\nabla_{\nu}\Phi)^2=& \ \nabla_{\mu}\nabla_{\nu}\Phi\nabla^{\mu}\nabla^{\nu}\Phi,\\[1ex]
	(\nabla_{\mu}\nabla_{\nu}\Phi)^3=& \ \nabla_{\mu}\nabla_{\nu}\Phi\nabla^{\nu}\nabla^{\lambda}\Phi\nabla_{\lambda}\nabla^{\mu}\Phi.
\end{align}
The $K$ and $G_i$ are free functions of the scalar field $\Phi$ and the kinetic term $X$. We represent each partial derivative with extra sub-index indicating the derivative's argument, i.e. $K_X=\p K/\p X$. 

The field equations derived from the action \eqref{action} is usually written as follows,
\begin{equation}
	\sum_{i=2}^{5}{\cal G}^i_{\mu\nu}=\kappa T_{\mu\nu},\quad \sum_{i=2}^{5}\left(\nabla^{\mu}J^i_\mu - P^i_\Phi\right)=0.
\end{equation}
The complete expressions for ${\cal G}^i_{\mu\nu}$, $J^i_\mu$ and $P^i_\phi$ can be found in the Appendix section of Ref.~\cite{Kobayashi:2011nu} and here we will only write their linearized version. Following the PN approximation scheme, we consider that each one of the scalar functions can be expanded as a power series about the background value $\varphi_0$. For any quantity $\xi(\varphi, X)$ or $\xi(\Phi, X)$, recalling that $\Phi = \varphi_0 + \varphi$, we write
\begin{equation}
	\xi(\varphi,X) \approx \xi_{(0,0)} + \xi_{(1,0)} \varphi + \xi_{(0,1)} X +... \, ,
\end{equation}
where $\xi_{(n,m)}$ are constants. It is not necessary to assume that $\xi$ is a true analytic function (i.e., that the power series above can be extended to infinity and it converges). It is assumed that, up to the necessary order, it is possible to approximate a function $\xi(\Phi,X)$ with the power series above.
Using the notation above, 
\begin{subequations} \label{Hseries}
\begin{align} 
	{\cal G}^2_{\mu\nu}& \approx -\frac{1}{2}(K_{(0,0)}+K_{(1,0)}\varphi)(\eta_{\mu\nu}+h_{\mu\nu}), \label{Hseries1}\\[1ex]
	{\cal G}^4_{\mu\nu}&\approx G_{4(0,0)}G_{\mu\nu}+ G_{4(1,0)}\left(\nabla^2\varphi\,\eta_{\mu\nu}-\p_\mu\p_\nu\varphi\right), \label{Hseries2} \\[1ex]
	J^2_\mu&\approx-K_{(0,1)}\p_\mu\varphi,\quad J^3_\mu \approx 2G_{3(1,0)}\p_\mu\varphi, \label{Hseries3}\\[1ex]
	P^2_\Phi&\approx 2K_{(2,0)}\varphi,\quad P^4_\Phi \approx G_{4(1,0)}R. \label{Hseries4}
\end{align}
\end{subequations}
The remaining terms vanish up to second PN order.

The procedure to solve the linearized field equations is the same as the one carried out in Section \ref{sec:st}. First, we use the trace equation to describe $R$ in terms of the scalar field and eliminate it from the other equations. For $\varphi$, one finds,
\begin{equation}
	\varphi=\frac{\kappa \,G_{4(1,0)}}{4\pi \, W}\, \int\frac{\rho(x',t)}{\x}\,e^{-m_{\varphi}\x}\,d^{3}x'\,,
\end{equation}
with,
\begin{equation}
W=G_{4(0,0)}\left(K_{(0,1)}-2G_{3(1,0)}\right) + 3G_{4(1,0)}^2,
\end{equation}
\begin{equation}
	m^2_\varphi=\frac{2}{W}\left(K_{(1,0)}G_{4(1,0)}-K_{(2,0)}G_{4(0,0)}\right).
\end{equation}
To solve the metric equations we generalize the gauge condition \eqref{gauge-hij} to,
\begin{equation}
	\p_kh^k{}_{i}+\frac{1}{2}\,\p_ih_{00}-\frac{1}{2}\,\p_ih^k{}_k=\frac{G_{4(1,0)}}{G_{4(0,0)}}\,\p_i\varphi.
\end{equation}
At the end, one will find,
\begin{align}
	\ba &= \frac{\kappa}{16\pi \, G_{4(0,0)}}\left(b_1+ b_2\,\frac{\varphi}{U} \right),\\[1ex]
	\bg &= \frac{\kappa}{16\pi \, G_{4(0,0)}}\left(b_3 - b_2\,\frac{\varphi}{U} \right), 
\end{align}
where we have defined,
\begin{align}
	b_1 &=2 + \frac{2K_{(1,0)}G_{4(1,0)}}{Wm_\varphi^2},\\[1ex]
	b_2 &= \dfrac{8\pi}{\kappa}\left(G_{4(1,0)}-\frac{K_{(1,0)}}{m^2_\varphi}\right),\\[1ex]
	b_3 &= 4-b_1.
\end{align}

Thus, the PN limit of Horndeski theories maintains the same pattern shown by the generalized Brans-Dicke theories. The contribution of the scalar field enters in the coefficients $\ba$ and $\bg$ with opposite signals, which leads to a constant $\gbg$ parameter. Once again, the analysis of the Newtonian limit will lead to different conditions for $\kappa$ in the cases of negligible or  large mass scalar field. Following the analysis of the previous section, one can extract the possible behavior for $\gbg$ and $\gamma$,
\begin{equation} \label{gbgGeneralH}
\!\!\!\!\!\!\gbg \!= \!
\begin{cases}
	\frac{\kappa}{4\pi\,G_{4(0,0)}} - 1\,,  \hspace{1.76cm} \mbox{in general},\\[.4cm]
	\! \gamma \!= \! 
	\begin{cases}
		\frac{W\,-\,G^2_{4(1,0)}}{W\,+\,G^2_{4(1,0)}} ,  & \! \! \mbox{if } m_\varphi^2 \ell^2  \sim O(1) ,  \\[.4cm]
		\frac{Wm^2_\varphi\,-\,K_{(1,0)}G_{4(1,0)}}{Wm^2_\varphi\,+\,K_{(1,0)}G_{4(1,0)}} ,  & \! \! \mbox{if }	e^{- m_\varphi \ell } \! \sim \! O(1).
	\end{cases} 
\end{cases}\!\!\!
\end{equation}
For the  commonly studied case $K_{(1,0)}=0$ \cite{Hohmann:2015kra, Hou:2017cjy},
\begin{equation}
\gbg \!= \!
\begin{cases}
	\frac{\kappa}{4\pi\,G_{4(0,0)}} - 1\,,  \hspace{0.6cm} \mbox{in general},\\[.4cm]
	\! \gamma \!= \! 
	\begin{cases}
		\frac{W\,-\,G^2_{4(1,0)}}{W\,+\,G^2_{4(1,0)}} ,  & \! \! \mbox{if } m_\varphi^2 \ell^2  \sim O(1) ,  \\[.4cm]
		1\, ,  & \! \! \mbox{if }	e^{- m_\varphi \ell } \! \sim \! O(1).
	\end{cases} 
\end{cases}\!\!\!
\end{equation}

It is important to stress that  $\gbg$  is not a spactial function, it is a constant. Consequently, $\gamma$ is a constant as well. Its value depends on the typical scale of the system $\ell$ in which the Newtonian limit is valid.

The  gravitational slip can be promptly computed and it is a spatial function given by
\begin{equation} \label{Hslip}
	\eta=\frac{\bg}{\ba}=\frac{b_3-b_2\varphi/U}{b_1+b_2\varphi/U}.
\end{equation}
As before, in the presence of a Newtonian limit in the studied system, the slip will coincide with $\gamma$ and $\gbg$. For systems in which a Newtonian limit is not assumed to hold (for instance, it may be relevant to consider that gravity in another galaxy does not have a Newtonian limit), then $\eta$ will be different from $\gbg$, while $\gamma$ will be ill defined.

The Horndeski Lagrangian reduces to the generalized Brans-Dicke theory discussed before by setting $K=2\omega X/\Phi - V$, $G_4=\Phi$ and $G_3=G_5=0$. Therefore, $K_{(1,0)}=-V_1$, $K_{(0,1)}=2\omega_0/\varphi_0$, $G_{4(0,0)}=\varphi_0$, $G_{4(1,0)}=1$, $W=3+2\omega_0$ and the  expressions \eqref{gbgGeneralH} and \eqref{Hslip} generalize the Brans-Dicke results of Section \ref{sec:st}.

References \cite{Hohmann:2015kra, Hou:2017cjy} compute several PPN parameters for Horndeski. For the negligible mass case, their $\gamma$ expression fully coincides with ours above. For the the large mass case, they find $\gamma=1$, which correspond to $K_{(1,0)}=0$. For the intermediary case, their expression for ``$\gamma(r)$'' is our expression for $\eta$ particularized for a point particle and with $K_{(1,0)}=0$ (the latter implies that $b_3=b_1=2$). As previously discussed, $\eta$ or ``$\gamma(r)$'' are not relevant in general for evaluating light trajectories (or imposing $\gamma$ bounds), their results suggest that light trajectories in Horndeski may behave differently in different regions of the considered system (see also Ref.~\cite{Renevey:2020tvr}). But, since $\gbg$ is a constant, and since a constant $\gbg$ fully parameterizes light trajectories (as discussed in Section \ref{sec:gbg}),  Horndeski theories have no novel spatial dependence on the trajectories of light. That is, the bending of light follows Eq.~\eqref{deltatheta}.  There is a hypothesis here that is not particularly restrictive, but it is relevant, namely the expansions in Eqs.~\eqref{Hseries} need to be valid. With this consideration, the constancy of $\gbg$ means that, if in Eq.~\eqref{deltatheta} a different dependence on the impact parameter $d$ is detected observationally, this would not violate GR alone, but the complete Horndeski theory as an action for gravity. 

For a more concrete test, valid even if the Newtonian limit is not assumed, one can consider systems with two Einstein rings. Such systems are rare, but they have already been observed \cite{Gavazzi:2008aq}. For simplicity, if the lens would be a point source (which is not the case in the cited observation, but we assume for simplicity),  from the knowledge of the value of $\delta \theta$ \eqref{deltatheta} for each one of the rings, one could simply divide these values and compare if the division would be compatible with the ratio between the radius of the two Einstein rings. For a more realistic approach, in which the lens is not a point mass, one needs to take into consideration the mass distribution, since each ring would be sensitive to a different mass. This would probably require assumptions on dark matter. Although this case of continuous mass lens introduces technical difficulties and will enlarge the error bars, in principle it is feasible, and it would be a strong test of gravity; since it has the potential of falsifying Horndeski gravity.

Another possibility is to test the constancy of $\gbg$, and hence the validity of Horndeski gravity, is to test light deflection or time delay within different subsystems. All of them should be compatible with the same $\gbg$ value.


\section{Conclusions}\label{sec:conclusion}

The differences between the gamma from PPN, its possible extensions and the gravitational slip are  subtle but with important consequences to the physical bounds, as here discussed. The importance of precise statements about them becomes higher with the crescent use of PPN and related formalisms in the context of extragalactic astronomy (e.g., \cite{Bolton:2006yz, Amendola:2012ys, Baker:2012zs,  Cao:2017nnq, Collett:2018gpf, Clifton:2018cef, Ferreira:2019xrr}). Here we considered three different gamma definitions, they are $\gamma$, $\bg$ and $\gbg$. The first one ($\gamma$) is a constant whose definition comes straight from the \wn~formalism, thus being the true ``$\gamma_{\mbox{\rm \tiny PPN}}$'' parameter. It is implicitly defined by the metric \eqref{ppnmetric}, hence it is only well defined in the presence of a well-posed Newtonian limit. The famous strong observational bounds are valid for $\gamma$. The second ($\bg$), which we name ``extended gamma'', is a straightforward and formal $\gamma$ extension: it does not depend on the Newtonian limit, nor it needs to be a constant, it single similarity with $\gamma$ is the slot it occupies in the metric \eqref{barmetric}. This latter gamma is useful for comparisons with cosmological parametrizations and as an intermediary step for other computations.  The third gamma ($\gbg$) is defined from the extended PPN metric \eqref{barmetric} and by the sum \eqref{gammaBG}. It was used by Berry and Gair \cite{Berry:2011pb} as an effective $\gamma$ to describe the bending of light and the time delay in the context of $f(R)$ gravity. As we have shown here, whenever $\gbg$ is a constant (and only in this case), it can fully parametrize light trajectories, even in the absence of a well defined Newtonian limit. At last, the slip parameter $\eta$ has several similar definitions in the literature (in essence it is a comparison, by means of a division, between two metric perturbations), we fixed here the one in eq.~\eqref{slipdeff}. It does not have the same physical implications of $\gamma$. When using the standard PPN metric \eqref{ppnmetric}, $\eta$ and all the gammas  coincide, but in general all these quantities are different among themselves, and with different physical implications. In particular, contrary to certain common belief (as discussed in Section \ref{sec:BDdiscussion}), $\eta$ can be far from the value of 1, its value can be varying significantly across space, and yet it is possible that the bending of light and the Shapiro time delay are the same of GR.

We have methodically applied the formalism to the well known cases of Brans-Dicke theory with a potential and to metric and Palatini $f(R)$ gravity, addressing some discrepancies with other references, but in agreement with Refs.~\cite{Berry:2011pb, Alsing:2011er, Clifton:2008jq}. We have explicitly shown here that the different approaches between \cite{Alsing:2011er} and reference \cite{Berry:2011pb}  are compatible, considering the phenomena associated to $\gamma$ (Section \ref{sec:BDintermediate}). In particular, $\gbg$ in these theories is always a constant.

Besides being in general wrong, we stress that there is no  computational advantage of using the slip as a parametrization for light deflection and the time delay, the computation of $\gbg$ is as easy as the computation of $\eta$. Also, in the absence of a Newtonian limit, one is not following a PPN approach, and the meaning of the PPN parameters can at most be valid under some conditions, that should be specified. We hope that this work stresses and clarifies this point. This is in  agreement with the approach of \cite{Alsing:2011er}, where they spelled out the relevant assumptions and introduced a modified PPN approach to deal with a theory without a proper Newtonian limit. Our approach here extends and it is closer to the approach of Ref.~\cite{Berry:2011pb}, since we do not modify $U$ and use $\gbg$. 

We have extended our results towards  Horndeski gravity. For this case, as expected, the slip is a space-time function \eqref{Hslip}, but although it is considerably more general than the BD case, $\gbg$ is also a constant in this case. This implies, as discussed in Section \ref{sec:gbg}, that $\gbg$ can be used to fully parametrize light trajectories, and that light trajectories behave exactly the same everywhere in the system, namely consistently with Eq.~\eqref{deltatheta}. This is valid even without assuming the existence of a Newtonian limit (but assuming that the Horndeski functions can approximately described by a power series, preserving the perturbative structure of post-Newtonian approaches). The constancy of $\gbg$ in a given system can in principle be tested, even if the Newtonian limit is not assumed. If it is, the situation is simpler,  there are two possible values for $\gamma$ \eqref{gbgGeneralH} and in principle one can use Newtonian gravity to model the mass distribution, besides detecting the light deflection locally. If Newtonian limit is not assumed, apart from doing specific model-dependent evaluations of the internal dynamics, one can use a general analysis by studying double Einstein ring systems which, in principle, allows for a verification on the constance of $\gbg$, as explained by the end of Section \ref{sec:horndeski}. 

To conclude, we notice that the analysis on the constance of $\gbg$ cannot consider subsystems that are cosmologically far apart. This since the complete syste cannot be properly described by a PN  asymptotically flat spacetime \eqref{ppnmetric}. Actually, for such setting, PPN Cosmology \cite{Sanghai:2016tbi, Clifton:2018cef} (or other similar approach) should be used to study the cosmological evolution of PPN parameters.


\begin{acknowledgments}
We thank Martin Makler for discussions on observational aspects and Oliver Piattella for  remarks on $f(R)$. JDT and DCR thank \textit{Conselho Nacional de Desenvolvimento Científico e Tecnológico} (CNPq, Brazil) and \textit{Fundação de Amparo à Pesquisa e Inovação do Espírito Santo} (FAPES, Brazil) for partial support.  Part of this work was developed while JDT were at UFES as a postdoc trough the  \textit{Profix} program, supported by FAPES and \textit{Coordenação de Aperfeiçoamento de Pessoal de Nível Superior} (CAPES, Brazil).
\end{acknowledgments}


\bibliography{bibdavi2019a,AllMyRefs}{}

\end{document}